\def\be{\begin{equation}}
\def\ee{\end{equation}}
\def\ba{\begin{array}}
\def\ea{\end{array}}

\def\Cb{{\Bbb C}}

\documentclass[pra,showpacs,twocolumn,amsmath]{revtex4}
\usepackage{epsfig}
\def\qed{\leavevmode\unskip\penalty9999 \hbox{}\nobreak\hfill
     \quad\hbox{\leavevmode  \hbox to.77778em{%
               \hfil\vrule   \vbox to.675em%
               {\hrule width.6em\vfil\hrule}\vrule\hfil}}
     \par\vskip3pt}

\newtheorem{theorem}{Theorem}
\newtheorem{corollary}{Corollary}

\input amssym.def
\begin{document}
\title{Lower Bound of Concurrence and Distillation for Arbitrary Dimensional Bipartite Quantum States}
\author{Ming-Jing Zhao$^{1}$}
\author{Xue-Na Zhu$^{2}$}
\author{Shao-Ming Fei$^{1,3}$}
\author{Xianqing Li-Jost$^{1}$}
\affiliation{$^1$Max-Planck-Institute for Mathematics in the Sciences, 04103
Leipzig, Germany\\
$^2$Department of Mathematics, School of Science, South
China University of Technology, Guangzhou 510640, China\\
$^3$School of Mathematical Sciences, Capital Normal University, Beijing
100048, China}

\begin{abstract}

We present a lower bound of concurrence for arbitrary dimensional
bipartite quantum states. This lower bound may be used to improve
all the known lower bounds of concurrence. Moreover, the lower bound
gives rise to an operational sufficient criterion of distillability
of quantum entanglement. The significance of our result is
illustrated by quantitative evaluation of entanglement for entangled
states that fail to be identified by the usual concurrence
estimation method, and by showing the distillability of mixed states
that can not be recognized by other distillability criteria.

\end{abstract}

\maketitle

\section{Introduction}

Quantum entanglement is a striking feature of quantum systems and
plays essential roles in some physical processes such as quantum
phase transitions in various interacting quantum many-body systems.
Quantum entangled states are the key physical resources in many
quantum information processing. An important issue in the theory of
quantum entanglement is to recognize and quantify the entanglement
for a given quantum state. Concurrence is one of the most important
measures of quantum entanglement \cite{W. K. Wootters, A. Uhlmann,
P. Rungta, S. Albeverio2001, D. A. Meyer, A. R. R. Carvalho}. For
mixed two-qubit states, an analytical formula of concurrence has
been derived \cite{W. K. Wootters}. For general high dimensional
case, due to the extremizations involved in the computation, only a
few analytic formulas of concurrence have been found for some
special symmetric states \cite{Terhal-Voll2000}.

To estimate the concurrence for general mixed states, efforts have
been made toward the lower bounds of concurrence. In Ref.
\cite{167902} a lower bound of concurrence that can be tightened by
numerical optimization over some parameters has been derived. In
Ref. \cite{K. Chen2005} analytic lower bounds of concurrence for any
dimensional mixed bipartite quantum states have been presented by
using the positive partial transposition (PPT) and realignment
separability criteria. These bounds are exact for some special
classes of states and can be used to detect many bound entangled
states. In Ref. \cite{H. P. Breuer} another lower bound of
concurrence for bipartite states has been presented from a new
separability criterion Ref. \cite{breuerprl}. A lower bound of
concurrence based on local uncertainty relations criterion is
derived in Ref. \cite{vicente}. This bound is further optimized in
Ref. \cite{zhang}. In Refs. \cite{edward,Y. Ou} the authors
presented lower bounds of concurrence for bipartite systems in terms
of a different approach, which has a close relationship with the
distillability of bipartite quantum states. In Ref. \cite{X. S. Li}
an explicit analytical lower bound of concurrence is obtained by
using positive maps, which is better than the ones in Refs. \cite{K.
Chen2005,H. P. Breuer} in detecting some quantum entanglement. The
lower bound for bipartite and multipartite systems derived in Ref.
\cite{F. Mintert2007,L. Aolita} are not only analytical but also experimentally
measurable in terms of local measurements on two copies of the quantum
state \cite{Y. Huang}. All these bounds give
rise to a good quantitative estimation of concurrence. They are usually
supplementary in detecting quantum entanglement.

In this paper we
present a lower bound of concurrence which may be used to improve
all these existing lower bounds of concurrence. Detailed examples
are given to show the improvement of the lower bounds of concurrence
in Refs. \cite{K. Chen2005,Y. Ou}.

Due to the influence of the environment, generally maximally
entangled pure states could evolve into mixed ones. To get the ideal
resource for quantum information processing, a possible approach is
distillation \cite{C. H. Bennett,M. Horodecki}. Nevertheless
operational necessary and sufficient criterion of distillability has
not been found yet. The reduction criterion \cite{M. Horodecki1999}
and the lower bound in Ref. \cite{Y. Ou} provide sufficient
conditions for distillability. It turns out that our lower bound of
concurrence also gives rise to a sufficient condition of
distillability that may improve the existing distillability
criteria.

This paper is organized as follows. In section II, we derive a new
lower bound of concurrence for arbitrary dimensional bipartite state
in terms of lower dimensional systems. In section III, we show
that our lower bound may be used to improve all the known lower
bounds of concurrence with detailed examples. In
section IV, as an application, we show that our lower bound is a
sufficient condition for distillability of quantum entanglement.
Conclusions are given in section V.

\section{Lower bound of concurrence}

Let $H_A$ and $H_B$ be $m$ and $n$ dimensional vector spaces
respectively. Any pure bipartite state $|\psi\rangle\in H_A \otimes
H_B$ has the form, \be\label{psi} |\psi\rangle=\sum_{i=1}^m
\sum_{j=1}^n a_{ij}|ij\rangle, \ee where $a_{ij}\in \Cb$,
$\sum_{ij}|a_{ij}|^2=1$, $\{|ij\rangle\}$ is the basis of $H_A
\otimes H_B$. The concurrence of $|\psi\rangle$ is given by \be
C(|\psi\rangle)=\sqrt{2(1-tr\rho_A^2)}, \ee where $\rho_A=tr_B
|\psi\rangle\langle \psi|$ is the reduced density matrix, $Tr_B$
stands for the partial trace over the  space $H_B$ \cite{W. K.
Wootters, A. Uhlmann, P. Rungta, S. Albeverio2001, D. A. Meyer, A.
R. R. Carvalho}. It can be further written as \cite{S.
Albeverio2001}, \be C(|\psi\rangle)=2\sqrt{\sum_{i<k}^m \sum_{j<l}^n
|a_{ij}a_{kl}-a_{il}a_{kj}|^2}. \ee The concurrence is extended to
mixed state $\rho$ by the convex roof $C(\rho)=\min \sum_i p_i
C(|\psi_i\rangle)$ for all possible ensemble realizations $\rho=\sum
p_i |\psi_i\rangle \langle \psi_i|$.

To evaluate $C(\rho)$, we project high dimensional states to ``lower
dimensional" ones. For a given bipartite $m\otimes n$ pure state
$|\psi\rangle$ in Eq. (\ref{psi}), we define an ``$s \otimes t$"
(unnormalized) pure state $|\psi\rangle_{s\otimes
t}=\sum_{i=i_1}^{i_s} \sum_{j=j_1}^{j_t} a_{ij}|ij\rangle$, $1<s<m$, $1<t<n$. We denote the concurrence of
$|\psi\rangle_{s\otimes t}$ by
$$C(|\psi\rangle_{s\otimes
t})=2\sqrt{\sum_{i_p<i_k}^s \sum_{j_q<j_l}^t
|a_{i_pj_q}a_{i_kj_l}-a_{i_pj_l}a_{i_kj_q}|^2}.
$$
There are ${m \choose s} \times {n \choose t}$ different $s \otimes t$
states $|\psi\rangle_{s\otimes t}$ for a given $|\psi\rangle$, where
$m \choose s$ and $n \choose t$ are the binomial
coefficients. For a mixed state $\rho$, we define its ``$s \otimes
t$" mixed states $\rho_{s\otimes t}=A\otimes B\rho A^\dagger\otimes
B^\dagger$, where $A=\sum_{i_p=1}^s |i_p\rangle\langle i_p|$ and
$B=\sum_{j_q=1}^t |j_q\rangle\langle j_q|$, $1<s<m$, $1<t<n$.
$\rho_{s\otimes t}$ has the following form,
\begin{widetext}
\begin{eqnarray}
&&\rho_{s\otimes t}=\left(
\begin{array}{cccccccccc}
\rho_{i_1j_1,i_1j_1} & \cdots & \rho_{i_1j_1,i_1j_t}  & \rho_{i_1j_1,i_2j_1} & \cdots \rho_{i_1j_1,i_2j_t} & \cdots &
\rho_{i_1j_1,i_sj_1} & \cdots & \rho_{i_1j_1,i_sj_t}\\
\cdots & \cdots &\cdots  &\cdots  &\cdots  &\cdots  &\cdots &\cdots &\cdots \\
\rho_{i_1j_t,i_1j_1} & \cdots & \rho_{i_1j_t,i_1j_t} & \rho_{i_1j_t,i_2j_1} & \cdots \rho_{i_1j_t,i_2j_t} & \cdots & \rho_{i_1j_t,i_sj_1} & \cdots & \rho_{i_1j_t,i_sj_t}\\
\rho_{i_2j_1,i_1j_1} & \cdots & \rho_{i_2j_1,i_1j_t} & \rho_{i_2j_1,i_2j_1} & \cdots \rho_{i_2j_1,i_2j_t} & \cdots & \rho_{i_2j_1,i_sj_1} & \cdots & \rho_{i_2j_1,i_sj_t}\\
\cdots & \cdots & \cdots &\cdots  &\cdots  &\cdots   &\cdots &\cdots &\cdots \\
\rho_{i_2j_t,i_1j_1} & \cdots & \rho_{i_2j_t,i_1j_t} & \rho_{i_2j_t,i_2j_1} & \cdots \rho_{i_2j_t,i_2j_t} & \cdots & \rho_{i_2j_t,i_sj_1} & \cdots & \rho_{i_2j_t,i_sj_t}\\
\cdots & \cdots & \cdots &\cdots  &\cdots  &\cdots  &\cdots &\cdots &\cdots \\
\rho_{i_sj_1,i_1j_1} & \cdots & \rho_{i_sj_1,i_1j_t} & \rho_{i_sj_1,i_2j_1} & \cdots \rho_{i_sj_1,i_2j_t} & \cdots & \rho_{i_sj_1,i_sj_1} & \cdots & \rho_{i_sj_1,i_sj_t}\\
\cdots & \cdots & \cdots &\cdots  &\cdots  &\cdots  &\cdots &\cdots &\cdots \\
\rho_{i_sj_t,i_1j_1} & \cdots & \rho_{i_sj_t,i_1j_t} & \rho_{i_sj_t,i_2j_1} & \cdots \rho_{i_sj_t,i_2j_t} & \cdots & \rho_{i_sj_t,i_sj_1} & \cdots & \rho_{i_sj_t,i_sj_t}
\end{array}
\right),
\end{eqnarray}
\end{widetext}
which are unnormalized bipartite $s \otimes t$ mixed states. The
concurrence of $\rho_{s\otimes t}$ is defined by $C(\rho_{s\otimes
t})\equiv\min \sum_i p_i C(|\psi_i\rangle_{s\otimes t})$, minimized
over all possible $s \otimes t$ pure state decompositions of
$\rho_{s\otimes t}=\sum_i p_i |\psi_i\rangle_{s\otimes t} \langle
\psi_i|$, with $\sum_i p_i=tr(\rho_{s\otimes t})$.

\begin{theorem}\label{th general lower bound}
For bipartite mixed state $\rho\in H_A \otimes H_B$,
\begin{eqnarray}\label{THM1}
C^2(\rho)\geq c_{st} \sum_{P_{st}} C^2(\rho_{s\otimes t})\equiv \tau_{s\otimes t}(\rho),
\end{eqnarray}
where $c_{st}=[{m-2 \choose s-2} \times {n-2 \choose t-2}]^{-1}$,
$\sum_{P_{st}}$ stands for summing over  all possible $s \otimes t$
mixed states, $\tau_{s\otimes t}(\rho)$ denotes the lower bound of
$C(\rho)$ with respect to the $s \otimes t$ subspace.
\end{theorem}

Proof. It is straightforward to prove that the concurrence of a pure
state $|\psi_i\rangle$ and  the concurrence of
$|\psi_i\rangle_{s\otimes t}$ associated to it have the following
relation,
\begin{eqnarray*}
C^2(|\psi_i\rangle)=c_{st}\sum_{P_{st}} C^2(|\psi_i\rangle_{s\otimes t}).
\end{eqnarray*}
Therefore for bipartite mixed state $\rho=\sum p_i |\psi_i\rangle \langle \psi_i|$, we have
\begin{eqnarray*}
\ba{rcl}
C(\rho)&=&\displaystyle\min \sum_i p_i C(|\psi_i\rangle)\\[1mm]
&=& \displaystyle\sqrt{c_{st}}\min \sum_i p_i \left(\sum_{P_{st}} C^2(|\psi_i\rangle_{s\otimes t})\right)^{\frac{1}{2}}\\[1mm]
&\geq&\displaystyle\sqrt{c_{st}} \min \left\{\sum_{P_{st}} \left[\sum_i p_i C(|\psi_i\rangle_{s\otimes t}) \right]^2\right\}^{\frac{1}{2}}\\[1mm]
&\geq& \displaystyle\sqrt{c_{st}}\left\{\sum_{P_{st}} \left[\min \sum_i p_i C(|\psi_i\rangle_{s\otimes t}) \right]^2\right\}^{\frac{1}{2}}\\[1mm]
&=&\displaystyle \sqrt{c_{st}} \left[\sum_{P_{st}} C^2(\rho_{s\otimes t})\right]^{\frac{1}{2}},
\ea
\end{eqnarray*}
where relation $(\sum_j (\sum_i x_{ij})^2 )^{\frac{1}{2}} \leq
\sum_i (\sum_j x_{ij}^2)^{\frac{1}{2}}$ has been used in the first
inequality, the first three minimizations run over all possible pure
state decompositions  of the mixed state $\rho$, while the last
minimization runs over all $s \otimes t$ pure state decompositions
of $\rho_{s\otimes t}=\sum_i p_i |\psi_i\rangle_{s\otimes t} \langle
\psi_i|$ associated to $\rho$. \qed

The lower bound of concurrence of $\rho$ in Eq. (\ref{THM1}) is
given by the concurrence of sub-matrix  $\rho_{s\otimes t}$.
Choosing different $s$ and $t$ would result in different lower
bounds. Generally we have
\begin{corollary}
\begin{eqnarray}
C^2(\rho)\geq \sum_{s=2}^m \sum_{t=2}^n p_{st} \tau_{s\otimes t}(\rho),
\end{eqnarray}
where $0\leq p_{st}\leq 1$, $\sum_{s=2}^m \sum_{t=2}^n p_{st}=1$.
\end{corollary}

Theorem \ref{th general lower bound} not only provides a lower bound
of concurrence, but also shows the  hierarchy among all the
concurrence of the lower dimensional bipartite mixed states
$\rho_{s\otimes t}$. The concurrence of $m\otimes n$ mixed state
$\rho$ is bounded by all the concurrence of the $(m-1)\otimes (n-1)$
mixed states associated to $\rho$, while the concurrence of the
$(m-1)\otimes (n-1)$ mixed state is bounded by that of all the
$(m-2)\otimes (n-2)$ related mixed states, and so on.

\section{Lower bound of concurrence from lower bounds}

Theorem \ref{th general lower bound} may be used to improve all the
known lower bounds of concurrence. Assume $g(\rho)$ is any lower
bound of concurrence, $C(\rho)\geq g(\rho)$.  Then for a given mixed
state $\rho$, the concurrence of the projected lower dimensional
mixed state $\rho_{s\otimes t}$ satisfies
\begin{eqnarray*}
C(\rho_{s\otimes t})
&&=tr(\rho_{s\otimes t})C((tr\rho_{s\otimes t})^{-1} \rho_{s\otimes t}) \\\nonumber
&&\geq tr(\rho_{s\otimes t}) g((tr\rho_{s\otimes t})^{-1}\rho_{s\otimes t}).
\end{eqnarray*}
Subsequently
\begin{eqnarray*}
C^2(\rho)&&\geq c_{st} \sum_{P_{st}} C^2(\rho_{s\otimes t}) \\\nonumber
&&\geq c_{st} \sum_{P_{st}} (tr\rho_{s\otimes t})^2g^2((tr\rho_{s\otimes t})^{-1}\rho_{s\otimes t}).
\end{eqnarray*}
Here if one chooses $\rho_{s\otimes t}$ to be the given mixed state
$\rho$ itself, the inequality  reduces to $C(\rho)\geq g(\rho)$
again. Generally the lower bound $g(\rho)$ may be improved if one
takes into account all the lower dimensional mixed states
$\rho_{s\otimes t}$. In this sense, we say the lower bound
$\tau_{s\otimes t}(\rho)$ may be used to improve all lower bound
$g(\rho)$. Let $g_i$, $i=1,...,k$, be a set of lower bounds of
concurrence. From Theorem  \ref{th general lower bound} and
Corollary 1 we have
\begin{theorem}
Let $g_i$ be a set of different lower bounds of concurrence. For any bipartite mixed state $\rho\in H_A \otimes H_B$,
\begin{widetext}
\begin{equation}\label{THM2}
C^2(\rho)\geq \sum_{s=2}^m \sum_{t=2}^n  \sum_{i=1}^k \sum_{P_{st}}  p_{st}\, c_{st}\, q_i\,
(tr\rho_{s\otimes t})^2 g^2_i\left((tr\rho_{s\otimes t})^{-1}\rho_{s\otimes t}\right),
\end{equation}
\end{widetext}
with
$0\leq p_{st}\leq 1$, $\sum_{s=2}^m \sum_{t=2}^n p_{st}=1$, $0\leq q_{i}\leq 1$, $\sum_{i=1}^k q_{i}=1$.
\end{theorem}

Therefore from any existing lower bounds of concurrence $g_i$,
$i=1,...,k$, one can get a new lower  bound (\ref{THM2}), which
satisfies $C(\rho)\geq~ \max\{g_1(\rho),...,g_k(\rho)\} $. Next we
give some explicit examples to demonstrate the improvement of a
given lower bound.

Let $|\psi\rangle=\sum_i \lambda_i |ii\rangle$ be an
arbitrary pure state in Schmidt  decomposition, and $f(\rho)$ a
real-valued and convex function such that
$f(|\psi\rangle\langle\psi|)\leq 2 \sum_{i<j} \lambda_i \lambda_j$
for any pure state $|\psi\rangle$. Then $f$ gives a lower bound of
concurrence for $m\otimes n$ ($m\geq n$) mixed states $\rho$,
$C(\rho) \geq \sqrt{\frac{2}{n(n-1)}} f(\rho)$ \cite{H. P. Breuer}.
In particular, concerning the partial transposition related lower
bound, we obtain the following lower bound.
\begin{corollary}
For any bipartite mixed state $\rho\in H_A \otimes H_B$,
\begin{eqnarray}\label{lower bound chen block}
C^2(\rho)&&\geq \frac{2c_{st}}{t(t-1)} \sum_{P_{st}} (||\rho_{s\otimes t}^{T_A}||- tr (\rho_{s\otimes t}))^2\\\nonumber
&&\equiv\kappa_{s\otimes t}(\rho),
\end{eqnarray}
where $t\leq s$ and $T_A$ stands for partial transpose with respect to the Hilbert space $H_A$.
\end{corollary}

Combining the result in Ref. \cite{K. Chen2005}, we can get another
lower bound that  improves the lower bound in Ref. \cite{K.
Chen2005}.
\begin{corollary}\label{cor lower bound chen block}
For any bipartite mixed state $\rho\in H_A \otimes H_B$,
\begin{equation}
\begin{array}{rcl}\label{lower bound chen block}
C^2(\rho)&\geq &
\frac{2c_{st}}{t(t-1)} {\displaystyle\sum_{P_{st}}}(\max (||\rho_{s\otimes t}^{T_A}||, ~||R(\rho_{s\otimes t})||)- tr (\rho_{s\otimes t}))^2\\[1mm]
&\equiv & \zeta_{s\otimes t}(\rho),
\end{array}
\end{equation}
where $t\leq s$, $R(\rho)$ stands for the realigned matrix of $\rho$.
\end{corollary}

The lower bound $\tau_{s\otimes t}(\rho)$ is better than the lower
bound in Ref. \cite{K. Chen2005}, since the latter is just a special
case of the lower bound $\zeta_{s\otimes t}(\rho)$ with $s=m$,
$t=n$,  $n\leq m$ in the Corollary \ref{cor lower bound chen block}.
As an example, we consider the mixed state
$\varrho_0=\frac{p}{3}(|00\rangle  + |11\rangle  + |22\rangle ) (
\langle 00| + \langle 11| + \langle 22| )+(1-p)|33\rangle \langle
33|$ with $0<p<1$. Its concurrence is $C(\varrho_0)=\frac{4}{3} p$.
The lower bound in Ref. \cite{K. Chen2005} gives
$C^2(\varrho_0)\geq\frac{2}{3} p^2$. Our lower bound shows that
$\tau_{3\otimes 3}(\varrho_0)=\frac{10}{9} p^2>\frac{2}{3} p^2$.

In the following we highlight the advantages of the lower bounds
$\tau_{s\otimes t}(\rho)$, $\kappa_{s\otimes t}(\rho)$ and
$\zeta_{s\otimes t}(\rho)$ by detailed analysis. First, the lower
bound $\tau_{s\otimes t}(\rho)$ is tight since the lower bound
$\tau_{2\otimes 2}(\rho)$ provides exact value for some mixed
states. Second, the lower bound $\tau_{s\otimes t}(\rho)$ is
strictly stronger than the lower bound $\tau(\rho)$ in Ref. \cite{Y.
Ou} because $\tau(\rho)$ is just a special case of our lower bound
with $s=t=2$. Since in two-qubit case, PPT implies separability, the
lower bound $\tau_{2\otimes 2}$ of any PPT entangled state $\rho$ is
zero. Therefore in this case $\tau_{2\otimes 2}$ can not detect any
PPT entanglement. However, the lower bound $\tau_{s\otimes t}$ and
$\zeta_{s\otimes t}$ can detect some PPT entangled states for $t>2$.

{\it Example 1.}
Let us consider the $4\otimes 4$ mixed state,
\begin{eqnarray*}
\varrho_1=p\sigma_\alpha +(1-p)|33\rangle \langle33|
\end{eqnarray*}
with $0<p<1$, $\sigma_\alpha = \frac{2}{7}|\psi^+\rangle \langle
\psi^+| +\frac{\alpha}{7}\sigma_+ +\frac{5-\alpha}{7}\sigma_-$,
$\sigma_+=\frac{1}{3}(|01\rangle \langle 01| + |12\rangle \langle
12| + |20\rangle \langle 20|)$, $\sigma_-=\frac{1}{3}(|10\rangle
\langle 10| + |21\rangle \langle 21| + |02\rangle \langle 02|)$,
$|\psi^+\rangle=\frac{1}{\sqrt{3}}(|00\rangle + |11\rangle +
|22\rangle)$. The state $\sigma_\alpha$ is separable for $2\leq
\alpha \leq 3$, bound (PPT) entangled for $3< \alpha \leq 4$, and
free entangled for $4< \alpha \leq 5$ \cite{P. Horodecki}. For PPT
entangled case, $\tau_{2\otimes 2}(\varrho_1)=0$  and
$C^2(\varrho_1)\geq 0$ from the lower bound in Ref. \cite{Y. Ou}.
While our lower bound gives $C^2(\varrho_1)\geq \tau_{3\otimes
3}(\varrho_1)\geq\zeta_{3\otimes
3}(\varrho_1)\geq\frac{p^2}{5292}(2\sqrt{3\alpha^2-15\alpha+19}-2)^2$,
which demonstrates the ability of our lower bounds in detecting PPT
entanglement.



{\it Example 2.}
For non-PPT (NPT) entangled state
\be\label{rho3}
\ba{rcl}
\varrho_2&=&\frac{p}{6}(|00\rangle \langle 00| + |01\rangle \langle 01| +|02\rangle \langle 02|+|10\rangle \langle 10|\\[1mm]
&&+|11\rangle \langle 11|+ |12\rangle \langle 12|)-\frac{p}{6}(|00\rangle \langle 12|+|01\rangle \langle 12|\\[1mm]
&&+|12\rangle \langle 00| +|12\rangle \langle 01| +|10\rangle \langle 11|+|11\rangle \langle 10|)\\[1mm]
&&+ \frac{1-p}{2}(|22\rangle\langle22| +|33\rangle\langle33|), \ea
\ee it can be verified that $\tau_{2\otimes 2}(\varrho_2)=0$ and
$\tau_{3 \otimes 3}(\varrho_2)\geq\kappa_{3 \otimes
3}(\varrho_2)=\frac{p^2}{54}>0$.  Hence $\tau_{3\otimes 3}$ provides
a better lower bound of concurrence for the NPT entangled state
$\varrho_2$.

We remark that the lower bound $\kappa_{2\otimes 2}(\rho)=0$ is
equivalent to $\tau_{2\otimes 2}(\rho)=0$.  That is,
$C(\rho_{2\otimes 2})=0$ is equivalent to $||\rho_{2\otimes
2}^{T_A}||- tr (\rho_{2\otimes 2})=0$ for every two-qubit state
$\rho_{2\otimes 2}$ of $\rho$. This is because two-qubit state
$\rho_{2\otimes 2}$ is PPT if and only if it is separable, i.e. zero
concurrence. Therefore $\kappa_{2\otimes 2}$ and $\tau_{2\otimes 2}$
could detect entanglement equivalently, up to a normalization. They
both present sufficient conditions for distillability of quantum
entanglement. Generally $\kappa_{2\otimes 2}$ can be more easily
computed than $\tau_{2\otimes 2}$.

\section{Lower bound of concurrence and distillation}

Distillation is an important protocol in improving the quantum
entanglement against the decoherence  due to noisy channels in
information processing. Theoretically
 an $m\otimes n$ mixed state $\rho$ is distillable if and only if there are some projectors $A$ and $B$
that map the $N$-copy state $\rho^{\otimes N}$ of $\rho$ to two-dimensional
ones such that the state $A\otimes B\rho^{\otimes N}A\otimes B$ is
entangled \cite{M. Horodecki}. Practically it is quite difficult to
judge whether an entangled mixed state is distillable in general.
From our lower bound of concurrence, we can derive the following
distillability criterion:

\begin{theorem}\label{distill}
 Any bipartite mixed state $\rho$ is distillable if $\tau_{2\otimes 3}(\rho^{\otimes N})>0$ for some positive integer $N$.
\end{theorem}

Proof. The conclusion in Ref. \cite{M. Horodecki} is also true for projectors $A$ and $B$ that
map the $N$-copy state $\rho^{\otimes N}$ to one two-dimensional space and one three-dimensional space respectively.
An $m\otimes n$ mixed state $\rho$ is distillable if $A\otimes B\rho^{\otimes N}A\otimes B$ is entangled.
This is due to that a mixed $2\otimes 3$ bipartite state $\rho$ is entangled if and only if it is distillable.
Therefore if $\tau_{2\otimes 3}(\rho^{\otimes N})>0$ for any
positive integer $N$, there will exist at least  one ``$2\otimes 3$
state" from $\rho^{\otimes N}$, which has nonzero concurrence. Hence
$\rho$ is distillable. \qed

The sufficient distillability condition in Ref. \cite{Y. Ou} says
that if $\tau_{2\otimes 2}(\rho^{\otimes N})>0$ for a certain
positive integer $N$, then $\rho$ is distillable.  Our sufficient
condition for distillability in Theorem \ref{distill} is stronger
than the one in Ref. \cite{Y. Ou}. The reason is that if
$\tau_{2\otimes 2}(\rho^{\otimes N})>0$, then $\tau_{2\otimes
3}(\rho^{\otimes N})>0$. But the converse is not true, since the
${2\otimes 2}$ submatrix mapped from the ${2\otimes 3}$ state may be
not included in the states in the bound $\tau_{2\otimes 2}$.

For example, let us consider the mixed state $\varrho_2$ in Eq.
(\ref{rho3}) in example 2. There we have  showed that
$\tau_{2\otimes 2}(\varrho_2)=0$. We now show that $\tau_{2\otimes
2}(\varrho_2^{\otimes N})=0$ for arbitrary positive integer $N$. To
do this we only need to prove that all the $2\otimes 2$ quantum
states from $\varrho_2^{\otimes N}$ are PPT for arbitrary positive
integer $N$. From the relations $\varrho_2^{\otimes
N}=\varrho_2^{\otimes N-1} \otimes \varrho_2$ and $(C\otimes
D)^{T_A}=C^{T_A}\otimes D^{T_A}$ for any operators $C$ and $D$
acting on $H_A \otimes H_B$, we have that all the principal minors
of order 2 from $\varrho_2$ and $\varrho_2^{\otimes N-1}$ are PPT,
therefore all the $2\otimes 2$ states from $\varrho_2^{\otimes N}$
are PPT and separable for any positive integer $N$. This implies
$\tau_{2\otimes 2}(\varrho_2^{\otimes N})=0$. Hence from the
sufficient condition of distillability in Ref. \cite{Y. Ou}, one
does not know whether the state $\varrho_2$ is distillable. In fact,
it is straightforward to verify that the state $\varrho_2$ does not
violate the reduction criterion \cite{M. Horodecki1999} either, so
the reduction criterion can neither detect its distillability.

But from the fact that $\tau_{2\otimes 3}(\varrho_2)=\tau_{3\otimes
3}(\varrho_2)>0$, we can assert  that state $\varrho_2$ is
distillable. Hence our sufficient condition of distillability is
strictly stronger than that in Ref. \cite{Y. Ou}.

\section{Conclusions}

In summary, we have proposed a novel method to construct hierarchy
lower bounds of concurrence for  arbitrary dimensional bipartite
mixed states, in terms of the concurrence of all the lower
dimensional mixed states extracted from the given mixed states. This
lower bound is no worse than any known lower bounds and may be used
to improve all the existing lower bounds of concurrence. The lower
bound has advantages in detecting PPT entanglement. Moreover, our
lower bound also gives rise to a sufficient condition for
distillability. This sufficient condition is complement to the
reduction criterion and is strictly stronger than the one in Ref.
\cite{Y. Ou}. Although our lower bound depends on the concurrence of
the lower dimensional mixed states and no analytical formulas of
concurrence for the lower dimensional mixed states are ready yet,
our lower bound still exhibits its excellence in entanglement
characterization and distillation.

\bigskip
\noindent{\bf Acknowledgments}\, M.J. Zhao would like to thank Y.C. Ou and B.B. Hua for valuable discussions.
 This work is supported by the NSFC Grants No. 10875081 and PHR201007107.


\begin{thebibliography}{18}

\bibitem{W. K. Wootters} W.K. Wootters, Phys. Rev. Lett. \textbf{80}, 2245 (1998).

\bibitem{A. Uhlmann} A. Uhlmann, Phys. Rev. A \textbf{62}, 032307 (2000).

\bibitem{P. Rungta} P. Rungta, V. Bu\v{z}ek, C.M. Caves, M. Hillery, and G.J. Milburn, Phys. Rev. A \textbf{64}, 042315 (2001).

\bibitem{S. Albeverio2001} S. Albeverio and S.M. Fei, J. Opt. B: Quantum Semiclass. Opt.
\textbf{3}, 223 (2001).

\bibitem{D. A. Meyer} D.A. Meyer and N.R. Wallach, J. Math. Phys. (N.Y.) \textbf{43},
4273 (2002).

\bibitem{A. R. R. Carvalho}  A.R.R. Carvalho, F. Mintert, and A. Buchleitner, Phys.
Rev. Lett. \textbf{93}, 230501 (2004).

\bibitem{Terhal-Voll2000}
P. Rungta and C.M. Caves, Phys Rev A \textbf{67}, 012307 (2003);
K. Chen, S. Albeverio, and S.M. Fei, Rep. Math. Phys. \textbf{58}, 325 (2006);
S.M. Fei, Z.X. Wang, and H. Zhao, Phys. Lett. A \textbf{329}, 414 (2004).

\bibitem{167902} F. Mintert, M. Ku\'{s}, and A. Buchleitner, Phys. Rev. Lett. \textbf{92}, 167902 (2004).

\bibitem{K. Chen2005} K. Chen, S. Albeverio, and S.M. Fei, Phys. Rev. Lett. \textbf{95}, 040504 (2005).

\bibitem{H. P. Breuer} H.P. Breuer, J. Phys. A: Math. Gen. \textbf{39}, 11847 (2006).

\bibitem{breuerprl} H.P. Breuer, Phys. Rev. Lett. \textbf{97}, 080501(2006).

\bibitem{vicente} J.I. de Vicente, Phys. Rev. A \textbf{75}, 052320 (2007).

\bibitem{zhang} C.J. Zhang, Y.S. Zhang, S. Zhang, and G.C. Guo. Phys. Rev. A \textbf{76}, 012334 (2007).

\bibitem{edward} E. Gerjuoy, Phys. Rev. A \textbf{67}, 052308 (2003).

\bibitem{Y. Ou} Y.C. Ou, H. Fan, and S.M. Fei, Phys. Rev. A \textbf{78}, 012311 (2008).

\bibitem{X. S. Li} X.S. Li, X.H. Gao, and S.M. Fei, Phys. Rev. A \textbf{83}, 034303 (2011).

\bibitem{F. Mintert2007} F. Mintert and A. Buchleitner, Phys. Rev. Lett. \textbf{98}, 140505 (2007).

\bibitem{L. Aolita} L. Aolita, A. Buchleitner, and F. Mintert, Phys. Rev. A \textbf{78}, 022308 (2008).

\bibitem{Y. Huang} Y.F. Huang, X.L. Niu, Y.X. Gong, J. Li, L. Peng, C.J. Zhang, Y.S. Zhang, and G.C. Guo, Phys. Rev. A \textbf{79}, 052338 (2009).

\bibitem{C. H. Bennett} C.H. Bennett, G. Brassard, S. Popescu, B. Schumacher, J.A. Smolin, and W.K. Wootters, Phys. Rev. Lett. \textbf{76}, 722 (1996).

\bibitem{M. Horodecki} M. Horodecki, P. Horodecki, and R. Horodecki, Phys. Rev. Lett. \textbf{80}, 5239 (1998).

\bibitem{M. Horodecki1999} M. Horodecki and P. Horodecki, Phys. Rev. A \textbf{59}, 4206 (1999).

\bibitem{P. Horodecki} P. Horodecki, M. Horodecki, and R. Horodecki, Phys. Rev. Lett. \textbf{82}, 1056 (1999).

\end{thebibliography}
\end{document}